\documentclass[final,5p,times,twocolumn]{elsarticle}
\usepackage{lineno}

\usepackage{amssymb}
\usepackage{amsmath}
\usepackage{siunitx}
\usepackage{hyperref}
\usepackage{graphicx}
\usepackage{microtype}
\usepackage{subcaption}
\usepackage{tabularx}
\usepackage{booktabs}

\newcommand{\mevee}{\ensuremath{\mathrm{MeV}_{\mathrm{ee}}}}
\newcommand{\kevee}{\ensuremath{\mathrm{keV}_{\mathrm{ee}}}}
\newcommand{\dd}{\mathrm{d}}
\newcommand{\nuc}[2]{\ensuremath{{}^{#1}\mathrm{#2}}}
\newcommand{\cdeut}{\ensuremath{\mathrm{C}_{6}\mathrm{D}_{6}}}

\hypersetup{
  colorlinks=true,
  citecolor=black,
  linkcolor=black,
  urlcolor=black,
  pdfauthor={A. Pérez de Rada Fiol et al.}
}

\journal{NIM A}

\begin{document}

\begin{frontmatter}

\title{Characterization of the fast neutron response of liquid scintillator detectors at NFS}

\author[ciemat]{A.~Pérez de Rada Fiol\corref{cor}}
\ead{alberto.rada@ciemat.es}
\cortext[cor]{Corresponding author at CIEMAT, Avenida Complutense 40, 28040 Madrid, Spain. Tel: +34 913466615.}

\author[ucm]{G.~del Río}
\author[ciemat]{T.~Martínez}
\author[ganil]{X.~Ledoux}
\author[ciemat]{V.~Alcayne}
\author[ciemat]{C.~Ballesteros}
\author[ciemat]{D.~Cano-Ott}
\author[ijclab]{I.~Matea}
\author[irfu]{P.~Miriot-Jaubert}
\author[ciemat]{J.~Plaza}
\author[ciemat]{A.~Sánchez-Caballero}
\author[irfu]{M.~Vandebrouck}
\author[ciemat]{D.~Villamarín}

\address[ciemat]{Centro de Investigaciones Energéticas, Medioambientales y Tecnológicas (CIEMAT), 28040 Madrid, Spain}
\address[ucm]{Universidad Complutense de Madrid (UCM), 28040 Madrid, Spain}
\address[ganil]{Grand Accélérateur National d'Ions Lourds (GANIL), 14076 Caen, France}
\address[ijclab]{IJCLab, Université Paris-Saclay, CNRS/IN2P3, 91405 Orsay, France}
\address[irfu]{Irfu, CEA, Université Paris-Saclay, 91191 Gif-sur-Yvette, France}

\begin{abstract}
The fast neutron response of MONSTER and sTED liquid scintillation detectors was investigated using time-of-flight measurements at the Neutrons For Science facility at GANIL-SPIRAL2. A continuous neutron beam produced by 40 MeV deuterons on a thick beryllium converter was used to obtain pulse height distributions for selected incident neutron energies. Digital pulse shape analysis was used to distinguish neutron-induced events and reduce pile-up effects. For MONSTER, the response and detection efficiency were studied up to 40 MeV using Geant4 with previously obtained light output functions and PHITS in SCINFUL mode. The comparison examines the increasing importance of carbon reaction channels and the divergence between the calculated efficiencies above 20 MeV. For sTED, light output functions for recoil deuterons and alpha particles were determined for the first time for this detector. The measured and calculated efficiencies were compared up to 20 MeV. The sTED efficiency shows a good overall agreement with the simulations. These results provide experimentally constrained detector response information for nuclear data measurements and identify limitations that must be considered when extending the simulations to higher neutron energies.
\end{abstract}


\end{frontmatter}


\section{Introduction}
\label{sec:introduction}
Organic liquid scintillators are widely used for fast neutron measurements due to the combination of a prompt response with a high detection efficiency over a broad energy range. Their pulse signals are produced mainly by charged particles generated in neutron interactions with the scintillator. Thus, the measured pulse height depends on the interaction channel, the energies of the charged products, their energy loss in the active volume, and the conversion of deposited energy into scintillation light. A neutron of fixed incident energy therefore produces a distribution of pulse heights rather than a single peak. Accurate knowledge of this distribution is needed to determine detection efficiencies, evaluate thresholds, and model the contribution of scattered neutrons in a complete experimental arrangement \cite{Schmidt2002,Dietze1982}.

For a neutron response expressed as a probability density in light output, the detection efficiency is the integral above the threshold, including the event selection conditions. This relation makes the light output function and the interaction model central to an efficiency calculation. Agreement with the most prominent recoil edge alone does not establish that the complete response, or its integral at a different threshold, is reproduced. The issue becomes particularly relevant when additional carbon reaction channels contribute to the response and when secondary charged particles can escape from the active volume. Detector characterization must therefore confront both the spectral distributions and the resulting efficiencies with measurements. In this work, we present the characterization of two different organic liquid scintillators.

On one hand, the MOdular Neutron time-of-flight SpectromeTER (MONSTER), was developed for neutron spectroscopy in beta decay experiments and in neutron reaction studies \cite{Garcia2012,Martinez2014,PerezdeRada2025b} through the time-of-flight (TOF) technique \cite{PerezdeRada2025a}. Its modules contain hydrogenous BC501A or EJ301 liquid scintillator. Earlier characterization campaigns established the light response information used in simulations at lower neutron energies, below 20 MeV. The treatment of neutron-induced alpha production on carbon was improved in Geant4 by incorporating the multistep breakup description and angular distribution information developed for NRESP \cite{Garcia2017}. Extending the experimental comparison into the tens-of-MeV region remains necessary because reaction channels producing protons and deuterons on carbon, together with other non-elastic processes, become increasingly important when using MONSTER in reaction studies \cite{Miriot2026}.

On the other hand, the segmented Total Energy Detector (sTED), was developed for neutron capture cross-section measurements at the high instantaneous flux of n\_TOF EAR2 \cite{Alcayne2024}. It consists of small \cdeut\ modules whose gamma ray response, timing properties, and suitability for the pulse height weighting technique have already been characterized. Knowledge of its fast neutron response is required to assess neutron-induced contributions to capture measurements and to investigate other uses of the detector. The replacement of hydrogen by deuterium changes the recoil kinematics and the relative contributions of the reaction channels.

The present work combines these two characterization problems using measurements at Neutrons For Science (NFS), where a continuous neutron spectrum and a long flight path allow response distributions to be selected by neutron time of flight \cite{Ledoux2021}. The results for MONSTER extend the efficiency study to 40 MeV and compare Geant4 with PHITS in SCINFUL mode. A direct application is the MONSTER efficiency used by \cite{Miriot2026} for neutron elastic scattering on natural cerium at 30.8 MeV. For sTED, the principal new results are its deuteron and alpha particle light output functions and its neutron detection efficiency up to 20 MeV.

\section{Experimental setup}
\label{sec:setup}
\subsection{Neutron beam}
The measurements were performed at the NFS facility at GANIL-SPIRAL2. A 40 MeV deuteron beam impinged on a thick beryllium converter, producing a continuous neutron spectrum through deuteron-induced reactions \cite{Ledoux2021}. The detectors under study were placed on the beam axis near the downstream end of the TOF hall, approximately 30 m from the converter. The beamline and collimation defined the incident neutron field at the irradiation position.

The accelerator radiofrequency supplied the time reference. Selection of one bunch out of approximately 440 reduced the nominal frequency of about 88 MHz to approximately 200 kHz, corresponding to a separation of about 5 microseconds between accepted bunches. This arrangement allowed neutron energy to be inferred from the flight time without overlap from consecutive bunches over the energy interval considered. Measurements with a carbon sample in the beam were used to identify resonance structures in neutron transmission for the TOF calibration. Direct irradiations without this sample provided the detector response distributions.

The operating current was reduced to approximately 0.1--0.2 microampere to limit the number of interactions per bunch. Nevertheless, the instantaneous intensity produced a significant probability of overlapping pulses in the detectors, making offline waveform treatment an essential part of the characterization.

A \nuc{238}{U}-based parallel-plate avalanche counter (PPAC) provided information on the neutron spectrum through the fission reaction. The neutron yield was expressed per unit energy, integrated deuteron charge and solid angle. Monitoring the beam in this way supplies the energy dependence needed for an efficiency determination, but the accuracy of the absolute scale also depends on the monitor calibration.

\subsection{Detector modules}
The MONSTER module used in this campaign was filled with EJ301 scintillator and coupled to a Hamamatsu R11833 photomultiplier. The sTED module was filled with deuterated benzene scintillator (EJ315) and coupled to a Hamamatsu R11265U-100 photomultiplier. The sTED measurement used a photomultiplier bias that was suitable for the lower pulse height region but did not provide a linear instrumental response over the full range accessible in the neutron beam. The efficiency results reported here are consequently restricted to incident neutron energies up to 20 MeV.

Signals were recorded with CIEMAT's Digital data AcquIsition SYstem (DAISY) \cite{Villamarin2023}, using full waveform acquisition employing digitizers with 14 bit amplitude resolution and a sampling rate of 1 GS/s. Stored waveform buffers made it possible to reprocess the data offline and treat overlapping events properly.

Light calibration measurements were carried out with standard gamma ray sources. The different Compton edges were used to establish the relationship between the recorded pulse amplitude and electron equivalent light output \cite{Kornilov2009}. The light resolution of both detectors was also obtained through these measurements.

\begin{figure}
\centering
\includegraphics[width=0.8\linewidth]{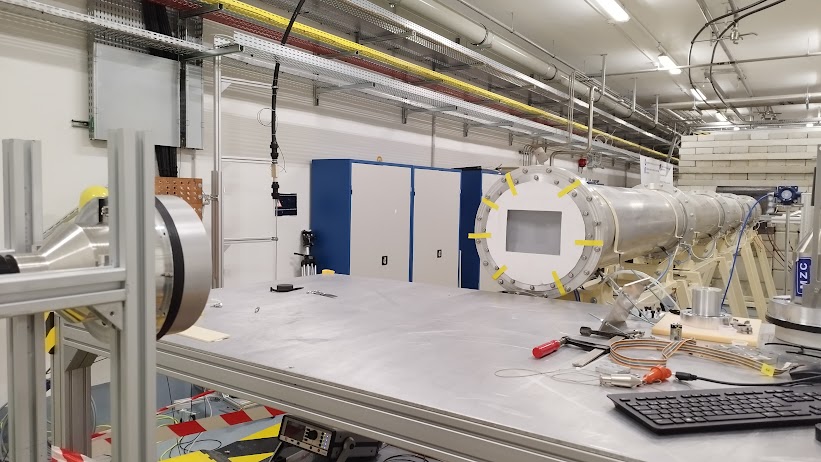}
\caption{MONSTER module positioned at the end of the NFS beamline.}
\label{fig:setup}
\end{figure}

A picture of the MONSTER module positioned at the end of the NFS beamline can be seen in Fig.~\ref{fig:setup}.

\section{Data analysis and Monte Carlo simulations}
\label{sec:analysis}
\subsection{Buffer reconstruction and neutron selection}
\label{sec:waveforms}
\begin{figure}
\centering
\includegraphics[width=0.9\linewidth]{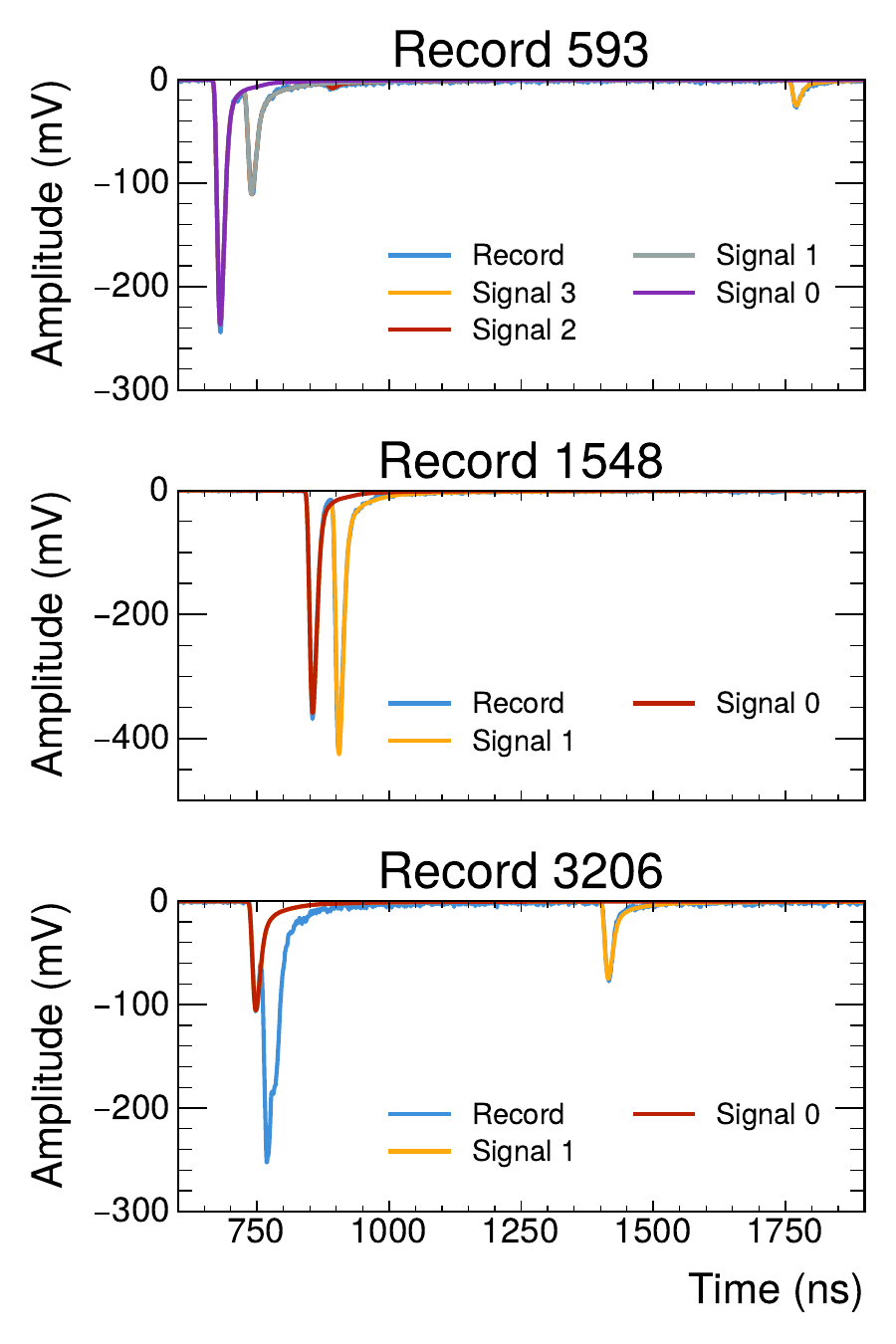}
\caption{The recorded buffer and the fitted individual signals are shown for three different cases.}
\label{fig:monster_pulses}
\end{figure}

At high instantaneous count rates, direct integration of a waveform can assign charge from a second interaction to the first pulse, distort its shape, or lose a small pulse on the tail of a larger one. These effects change both the pulse height distribution and the number of events. To address this problem, the analysis used average neutron and gamma ray pulse shapes as templates, following the true pulse shape method \cite{Guerrero2008}. Templates for neutron and gamma ray events were constructed by averaging large samples of isolated pulses after alignment and normalization. Comparison of the fit residuals for alternative pulse classes provides a neutron/gamma ray discrimination criterion. Fitting several signals allows partially overlapping signals to be separated when the waveform contains sufficient timing information. It cannot recover information lost through complete saturation or guarantee separation of arbitrarily close interactions. This can be seen in Fig.~\ref{fig:monster_pulses}, where different buffers recorded with the MONSTER module are reconstructed with this method.

\begin{figure}
\centering
\includegraphics[width=\linewidth]{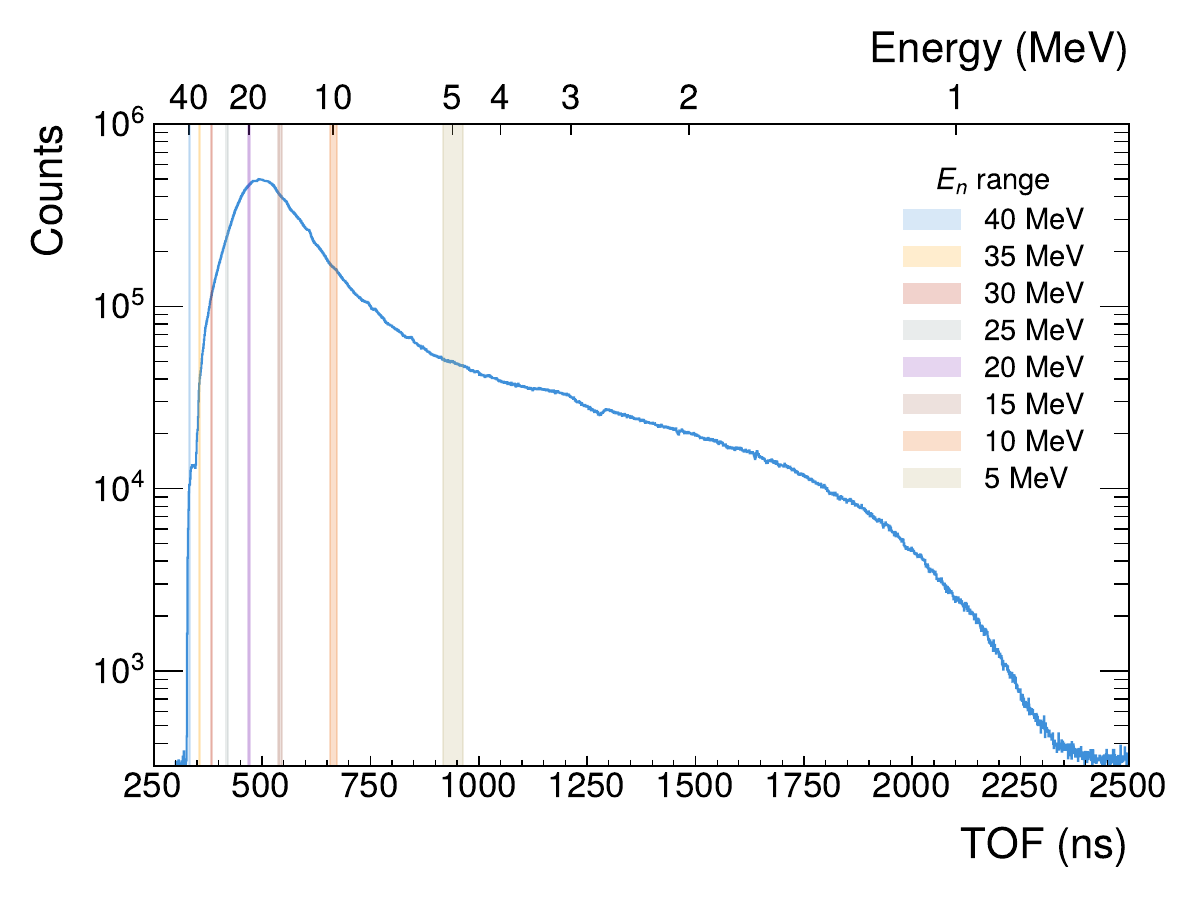}
\caption{Measured MONSTER TOF spectrum and selected windows to obtain quasi-monoenergetic pulse height distributions.}
\label{fig:monster_tof}
\end{figure}

\subsection{Light output functions}
\label{sec:light}
Neutron response distributions were obtained directly by selecting narrow TOF windows, as can be seen in Fig.~\ref{fig:monster_tof}, and histogramming the corresponding light output. The energy width was chosen so that broadening of the relevant recoil edge by the incident energy interval remained small compared with the detector light resolution \cite{Ohrn2008}. The selected distributions are consequently quasi-monoenergetic.

For MONSTER, the Geant4 simulations used previously obtained light output functions. The present study assesses their use in the extended neutron energy range rather than introducing a new fit of MONSTER's proton light response. The earlier characterization and the Geant4 treatment of organic scintillators provide the methodological basis \cite{Martinez2014,Garcia2017}.

For sTED, effective light functions were determined for deuterons and alpha particles \cite{DelRio2026}. The maximum recoil energy from elastic neutron scattering on a nucleus of mass ratio $A$ is, in the non-relativistic limit,
\begin{equation}
 T_{R,\max}=\frac{4A}{(1+A)^2}E_n.
 \label{eq:recoil}
\end{equation}
For deuterium this gives $T_{d,\max}=8E_n/9$. Recoil edges in the TOF-selected spectra were associated with these energies to obtain pairs of deuteron energy and light output. The angular dependence of neutron--deuteron scattering affects the shape preceding the edge and must be included in the response calculation \cite{Schwarz1983}.

The light functions were parameterized using an integrated Birks expression \cite{Birks1951},
\begin{equation}
 L_j(T)=S\int_0^T\frac{\dd E}{1+kB_j\,s_j(E)},
 \qquad s_j(E)=\frac{1}{\rho}\left(\frac{\dd E}{\dd x}\right)_j,
 \label{eq:birks}
\end{equation}
where $s_j$ is the positive mass stopping power for particle species $j$, $S$ is the light scale parameter, and $kB_j$ is an effective quenching parameter. Stopping powers were obtained with SRIM \cite{Ziegler2010}.

The alpha particle function was constrained using structures associated with the ground state channel of $\nuc{12}{C}(n,\alpha)\nuc{9}{Be}$. The reaction kinematics and the simulated charged particle spectra were used to relate the observed light structure to alpha energy. A nearby neutron energy window below the alpha production threshold provided an estimate of the underlying deuteron contribution. The strongly quenched beryllium recoil was neglected in the alpha light extraction as an approximation of the analysis. Only a limited alpha energy interval was available. In the regime $kB_\alpha s_\alpha\gg1$, Eq.~\eqref{eq:birks} depends predominantly on the ratio $S/kB_\alpha$. The two parameters are then strongly correlated. The alpha fit therefore fixed $S$ to the value obtained for deuterons and varied $kB_\alpha$. A proton light function was inferred, when needed for interpretation, through the approximate scaling relation $L_d(T_d)=2L_p(T_d/2)$ used in the NRESP approach \cite{Dietze1982}.

\begin{figure}
\centering
\includegraphics[width=0.9\linewidth]{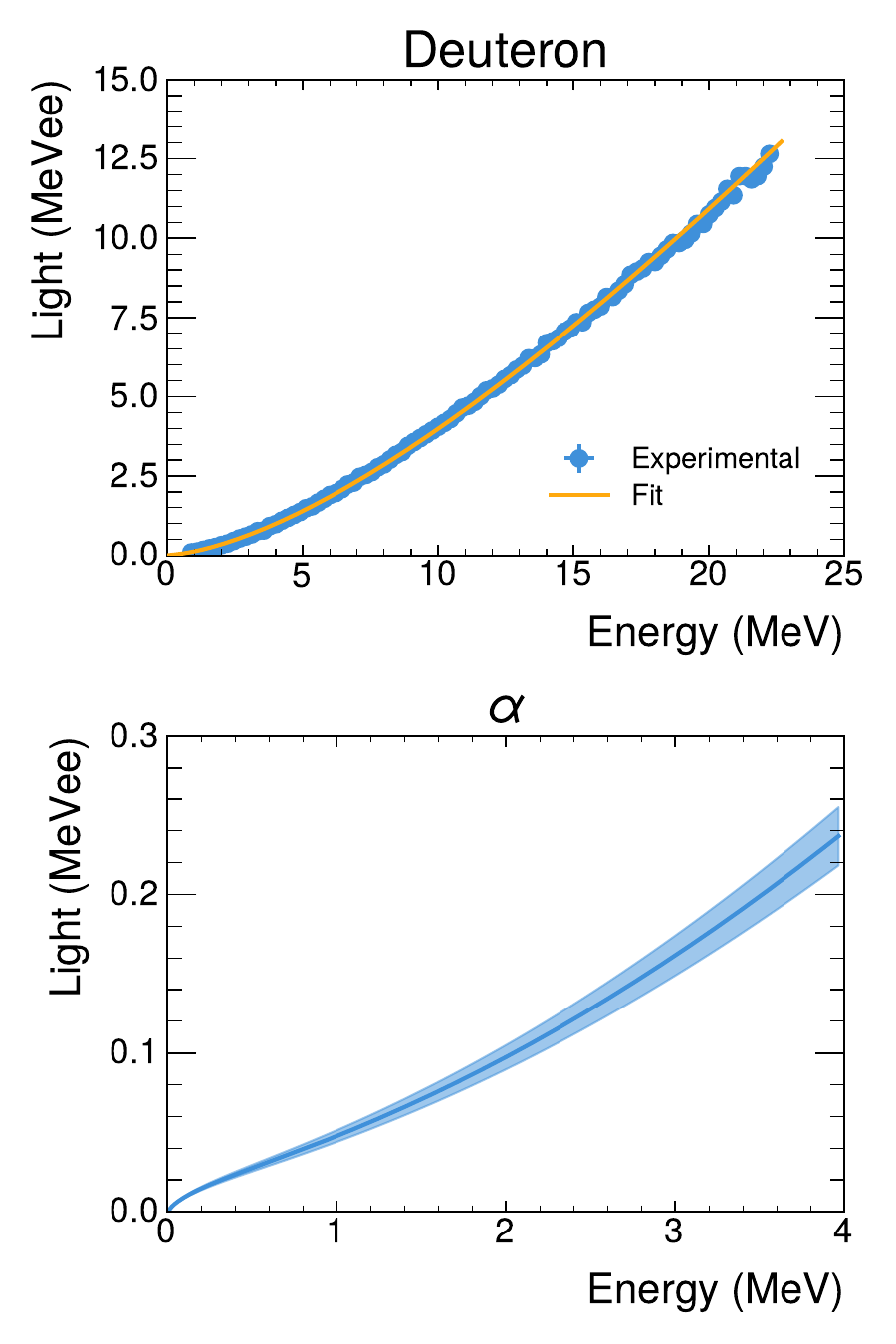}
\caption{Light output functions from the sTED analysis.}
\label{fig:light}
\end{figure}

The sTED analysis yielded the effective Birks parameters listed in Table~\ref{tab:birks}. The corresponding light functions are shown in Fig.~\ref{fig:light}. The deuteron fit gave $S=1.31\pm0.02$ \mevee/MeV and $kB_d=16.3\pm0.4$ mg cm$^{-2}$ MeV$^{-1}$, with a reduced $\chi^2$ of 1.167. With the light scale fixed to the deuteron result, the alpha fit gave $kB_\alpha=14\pm2$ mg cm$^{-2}$ MeV$^{-1}$ and a reduced $\chi^2$ of 1.073. These results establish particle dependent light response information for the sTED module.

\begin{table}
\centering
\caption{sTED light function parameters.}
\label{tab:birks}
\small
\begin{tabular}{@{}lcc@{}}
\toprule
Parameter & Deuteron & Alpha \\
\midrule
$S$ (\mevee/MeV) & $1.31\pm0.02$ & Fixed to $S_d$ \\
$kB$ (mg cm$^{-2}$ MeV$^{-1}$) & $16.3\pm0.4$ & $14\pm2$ \\
$\chi^2/{\rm ndf}$ & 1.167 & 1.073 \\
\bottomrule
\end{tabular}
\end{table}

\subsection{Monte Carlo simulations}
\label{sec:simulation}
The detector response simulations were performed with Geant4 \cite{Agostinelli2003,Allison2016}. For both detectors, the geometry included the active liquid, aluminium housing and support, optical elements, and the main photomultiplier components. Neutron simulations used monoenergetic primaries incident along the beam direction in a circular source plane with a radius of 13 cm to represent the beam shape at the irradiation position. The Shielding physics list with the ParticleHP \cite{Mendoza2011} package, together with the organic scintillator extensions required to describe the $\nuc{12}{C}(n,n'3\alpha)$ and $\nuc{12}{C}(n,\alpha)\nuc{9}{Be}$ reactions \cite{Garcia2017}, were used. Energy deposits from the individual charged particles were converted to light using the corresponding functions. For a continuously slowing particle, the light contribution of a track follows the difference between the light function values at its initial and final kinetic energies. The contributions of all relevant tracks are then summed for each incident neutron. This treatment accounts for incomplete energy deposition when a charged particle leaves the active volume. It also avoids applying a single light function to an event containing several particle species. The calculated event light was subsequently broadened according to the experimental resolution.

For MONSTER, an additional simulation was performed with PHITS \cite{Sato2018} in the mode that incorporates the SCINFUL approach for organic liquid scintillators \cite{Satoh2022}. The comparison shows different descriptions of neutron interactions and scintillation response, particularly beyond the energy range in which the established Geant4 carbon treatment had been validated.

\section{Results and discussion}
\label{sec:results}

\begin{figure*}
\centering
\includegraphics[width=0.8\textwidth]{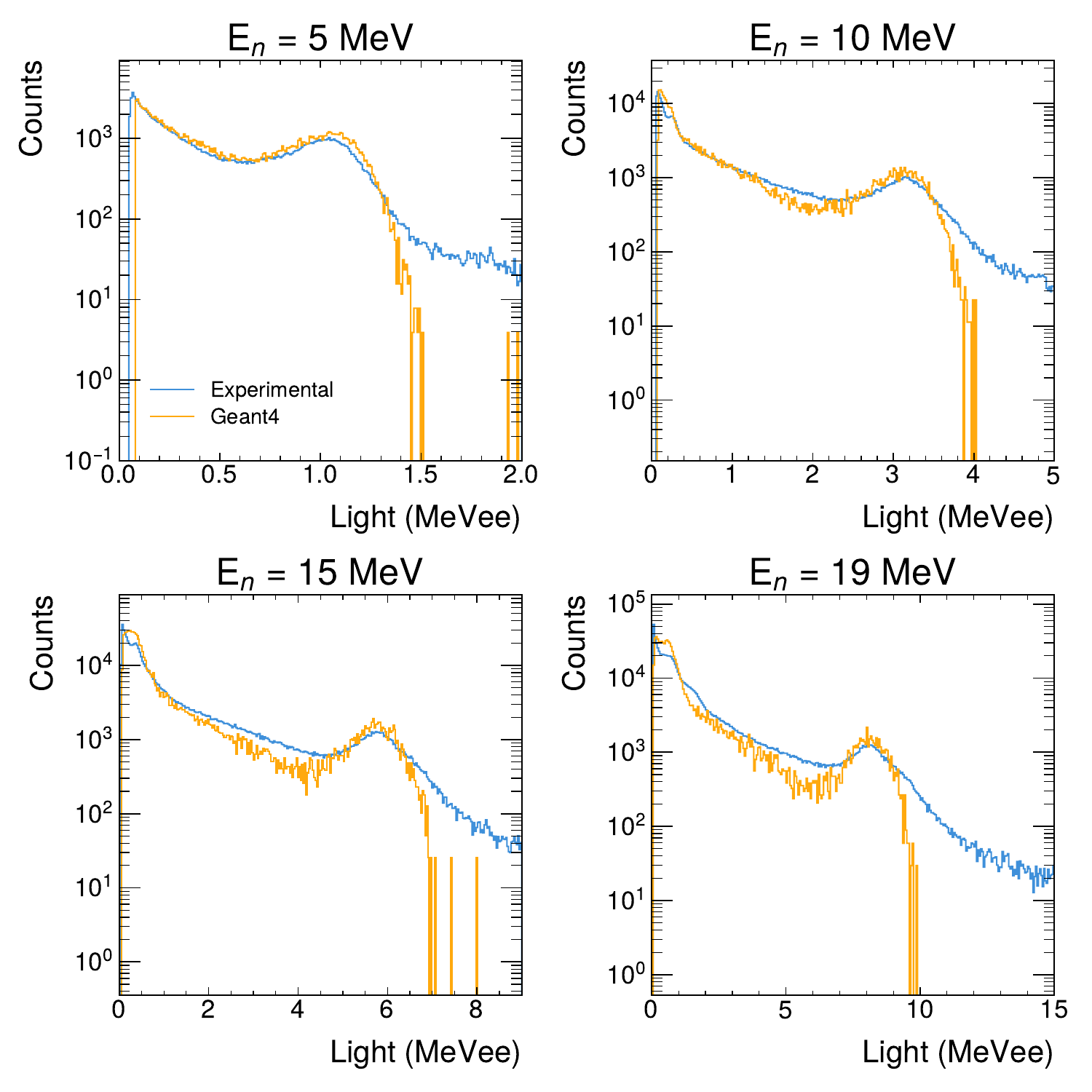}
\caption{Measured sTED pulse height distributions and the corresponding Geant4 calculations.}
\label{fig:stedresponse}
\end{figure*}

\begin{figure*}
\centering
\includegraphics[width=0.5\textwidth]{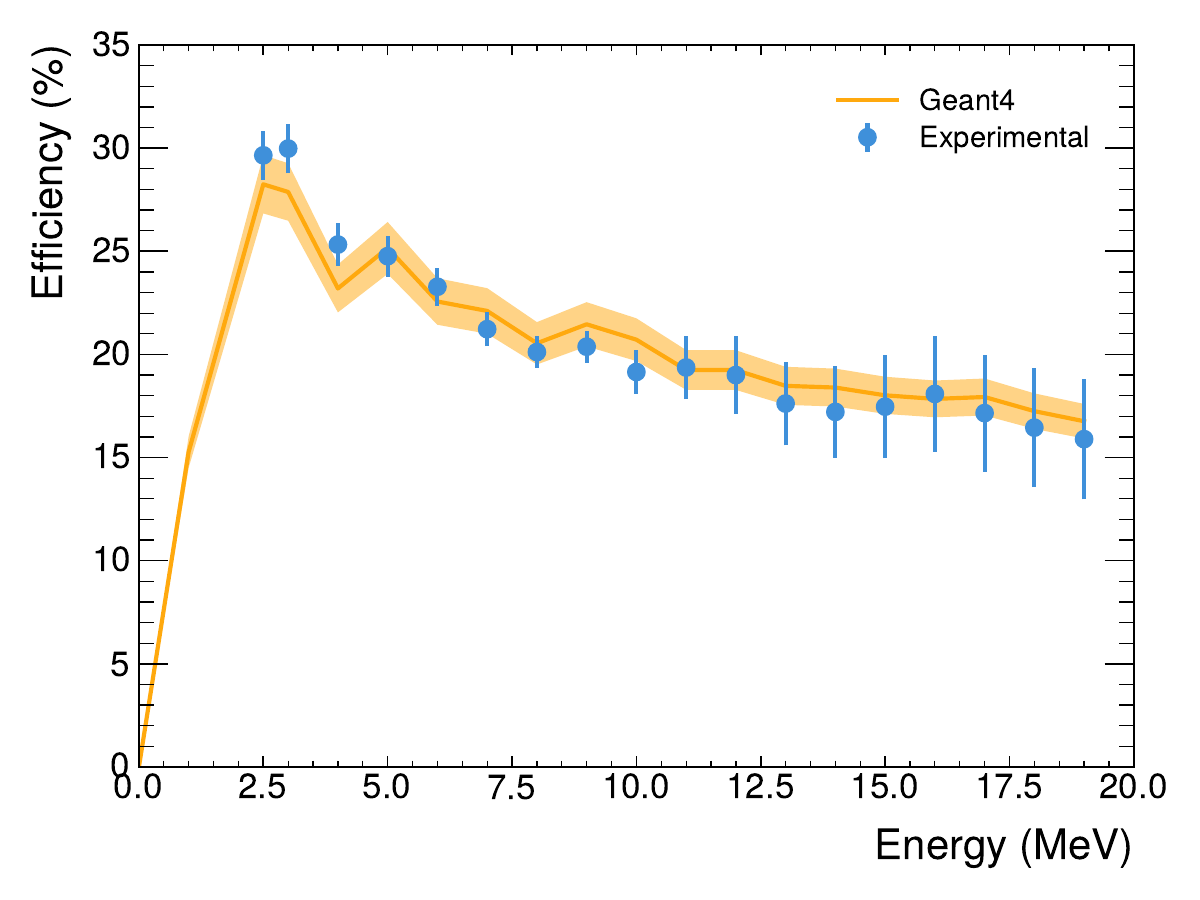}
\caption{sTED neutron detection efficiency.}
\label{fig:stedeff}
\end{figure*}

\begin{figure*}
\centering
\includegraphics[width=0.8\textwidth]{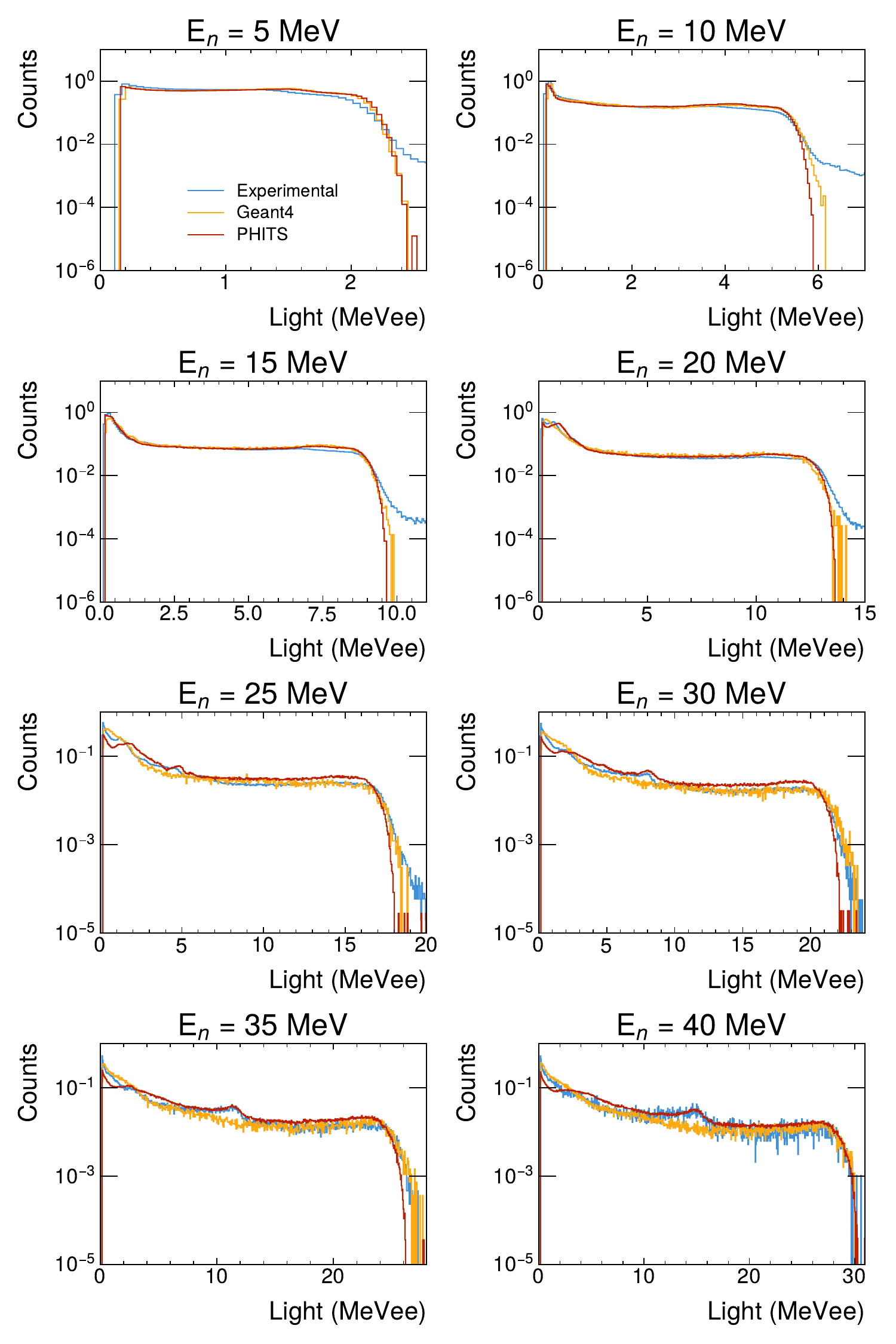}
\caption{Measured MONSTER pulse height distributions and the corresponding Geant4 and PHITS calculations.}
\label{fig:monsterresponse}
\end{figure*}

\subsection{sTED neutron response}
\label{sec:stedresponse}
Fig.~\ref{fig:stedresponse} compares measured and simulated sTED pulse height distributions at incident neutron energies of 5, 10, 15, and 19 MeV. The simulations reproduce the main evolution of the deuteron recoil structure with energy. The waveform treatment improves the correspondence near that structure by reducing the redistribution of counts caused by overlapping signals. Nevertheless, residual differences remain both at intermediate light output and beyond the main recoil edge.

The carbon channels are particularly important for interpreting the lower light structures once alpha production is energetically accessible. The nonlinear relation between particle energy and scintillation light means that different partitions of a fixed available energy among several alpha particles do not produce the same pulse height. A response model must therefore describe the charged particle final states, not only the total reaction cross section. This was the motivation for the Geant4 carbon developments \cite{Garcia2017}. In the comparison, some of the calculated alpha-associated structures remain more pronounced than the experimental ones. A further contribution can arise from neutron-induced deuteron breakup,
\begin{equation}
 n+d\longrightarrow n+n+p.
 \label{eq:breakup}
\end{equation}
Its threshold is approximately 3.34 MeV. The analysis identified incomplete production of the charged residual in the simulation as a possible explanation for part of the missing proton-induced light. The kinematics of the breakup reaction allow protons to populate a broad interval of light output. However, a pure phase-space construction cannot determine the detailed spectrum or its absolute contribution. Measurements by \cite{Pauletta1975} show why the reaction dynamics must be considered in addition to kinematic accessibility. The approximate proton light function inferred from the deuteron result is therefore useful for locating a possible contribution.

Likewise, the tails above the dominant recoil structure do not have a single origin. Residual overlapping events, a finite TOF response, particle interactions in surrounding materials, and backgrounds selected by the same time window can contribute. Multiple scattering inside the detector also distinguishes the full response from an ideal single-collision recoil edge.

The sTED efficiency obtained with an 80 \kevee\ threshold is shown in Fig.~\ref{fig:stedeff}. The curve reaches approximately 30\% around 2.5--3 MeV and then decreases to roughly 17\% near the upper end of the measured energy range. The increase near the threshold reflects the growing fraction of neutron interactions that produce sufficient light to surpass it. At higher neutron energy, the interaction probability and the distribution of secondary particle energy losses dominate the trend. Escape of charged products from the small active volume can reduce the light collected from an interaction.

The measured and simulated efficiencies agree more closely over most of the interval than some of the detailed spectral features in Fig.~\ref{fig:stedresponse}. This behaviour is possible because the efficiency integrates a broad range of light output. Differences in local structures can partly compensate in the integral, or remain above a threshold sufficiently low to count events despite their redistribution in light. The non-linearity observed at higher pulse heights prevents a reliable extension of the present sTED efficiency result beyond 20 MeV. Measurements at reduced PMT gain would allow that instrumental limitation to be separated from the nuclear physics description.

\subsection{MONSTER neutron response up to 40 MeV}
\label{sec:monsterresponse}
The MONSTER analysis extends to incident neutron energies of 40 MeV. Representative response distributions are shown in Fig.~\ref{fig:monsterresponse}. Up to 20 MeV the Geant4 and PHITS simulations predict similar main spectral distributions, both in agreement with the experimental spectra. Above 20 MeV, the two simulations start to differ more visibly, and neither calculation reproduces every feature of the measured spectra. At these incident energies, carbon reactions producing protons and deuterons contribute alongside elastic proton recoils and alpha production channels. Both their production probabilities and the light generated by their charged products affect the response. The comparison provides an experimental test of both simulation codes.

\begin{figure}
\centering
\includegraphics[width=\linewidth]{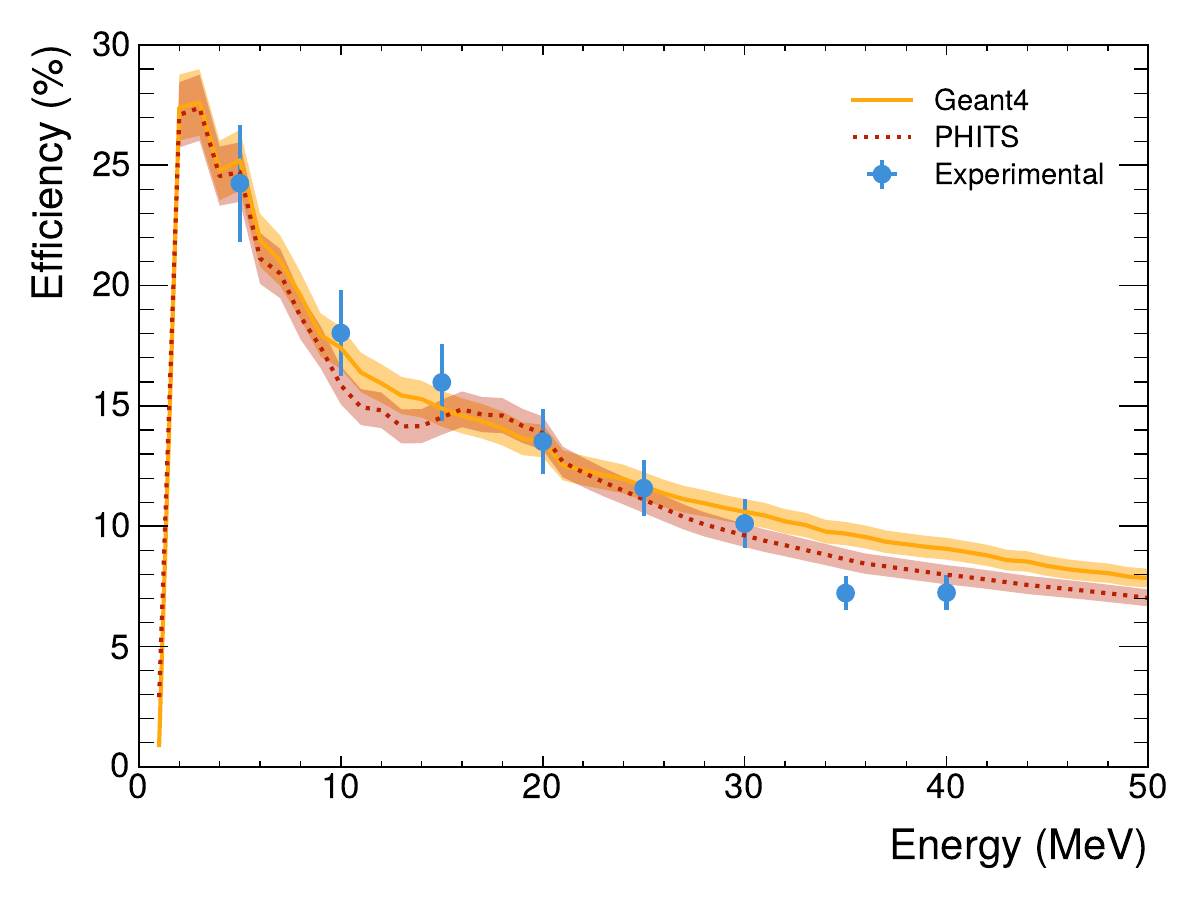}
\caption{MONSTER detection efficiency.}
\label{fig:monstereff}
\end{figure}

Fig.~\ref{fig:monstereff} presents the comparison of the measured MONSTER efficiency, obtained with a 180 \kevee\ threshold, and the Geant4 and PHITS simulated curves. The agreement between the two simulation codes and the measured data is at approximately the 5\% level below 20 MeV. Above 20 MeV the calculated efficiencies separate, with Geant4 generally predicting a larger-than-measured value. The highest energy experimental values lie below both calculations, although the PHITS one remains in reasonable agreement. These observations show why an extension of the ParticleHP Geant4 package is needed for higher energy neutrons.

The result of this characterization has already been used in \cite{Miriot2026} to obtain the neutron elastic scattering cross section on natural cerium at 30.8 MeV.

\section{Conclusions}
\label{sec:conclusions}

This work extends the neutron detection efficiency characterization of MONSTER to 40 MeV and provides the first determination of deuteron and alpha particle light output functions for sTED, together with its neutron detection efficiency up to 20 MeV. TOF measurements at NFS provided individual neutron response distributions for both detectors. Offline waveform reconstruction was essential for extracting these distributions in the presence of overlapping pulses.

For MONSTER, the measurements provide an experimental reference for Geant4 simulations using previously obtained light output functions and for PHITS simulations in SCINFUL mode. The two codes give similar efficiencies below 20 MeV, while their increasing differences at higher energies demonstrate the importance of the experimental characterization over the extended range. The resulting efficiency information has already been used in the determination of neutron elastic scattering cross section on natural cerium at 30.8 MeV \cite{Miriot2026}. This application demonstrates the direct contribution of the present work to quantitative neutron reaction measurements.

For sTED, the new light output functions establish the response to recoil deuterons and alpha particles. Their implementation in Geant4 reproduces the overall measured neutron efficiency dependence up to 20 MeV. The comparison also identifies differences in individual spectral structures, providing experimental constraints for improving the calculated response beyond its integrated efficiency. The upper energy limit is imposed by instrumental non-linearity under the measurement conditions. The light output functions and efficiency benchmark obtained here provide the basis for calculating fast neutron contributions to the response of sTED in neutron-induced reaction experiments.

\section*{CRediT authorship contribution statement}

\textbf{A.~Pérez de Rada Fiol:} Data curation, Formal analysis, Investigation, Methodology, Software, Supervision, Validation, Visualization, Writing – original draft.
\textbf{G.~del Río:} Formal analysis, Investigation, Writing – original draft.
\textbf{T.~Martínez:} Conceptualization, Funding acquisition, Investigation, Project administration, Supervision, Writing – review \& editing.
\textbf{X.~Ledoux:} Investigation, Resources.
\textbf{V.~Alcayne:} Investigation.
\textbf{C.~Ballesteros:} Investigation.
\textbf{D.~Cano-Ott:} Funding acquisition, Investigation, Project administration.
\textbf{I.~Matea:} Investigation.
\textbf{P.~Miriot-Jaubert:} Investigation.
\textbf{J.~Plaza:} Investigation, Software, Writing – review \& editing.
\textbf{A.~Sánchez-Caballero:} Investigation.
\textbf{M.~Vandebrouck:} Investigation.
\textbf{D.~Villamarín:} Investigation, Software.

\section*{Declaration of competing interest}
The authors declare no competing interests.

\section*{Acknowledgments}

This project has received funding from the Euratom research and training programme 2021–2025 under grant agreement No 101164596 (APRENDE). Views and opinions expressed are however those of the author(s) only and do not necessarily reflect those of the European Union or the European Atomic Energy Community (‘EC-Euratom’). Neither the European Union nor the granting authority can be held responsible for them.

This work was supported in part by project PID2024-158174OB-I00 funded by  MCIN/AEI/10.13039/501100011033/FEDER, UE.

The authors thank the GANIL-SPIRAL2/NFS staff for providing the neutron beam and supporting the measurements.

\section*{Data availability}

Data will be made available upon reasonable request.

\section*{Declaration of generative AI and AI-assisted technologies in the manuscript preparation process}

During the preparation of this work, the author(s) used ChatGPT for writing assistance. The author(s) reviewed and edited the output as needed and take full responsibility for the content of the published article.

\bibliographystyle{elsarticle-num}
\bibliography{references}

\end{document}